\documentclass[12pt]{article}

\usepackage{fixltx2e}

 \usepackage[doublespacing, nodisplayskipstretch]{setspace}

\usepackage[english, colorlinks=true, citecolor=ForestGreen]{hyperref}

\usepackage[english]{varioref}

\usepackage{amsmath}

\usepackage{amsfonts}

\usepackage{cmap}

\usepackage{mathtools}

\usepackage{amssymb}

\usepackage{icomma}

\usepackage{latexsym}

\usepackage[T2A]{fontenc}			

\usepackage[utf8]{inputenc}	

\def\keywordfont{\fontsize{9}{10}\selectfont\leftskip1pc\rightskip5pc plus1fill}%
\usepackage[T2A]{fontenc}			
\usepackage{icomma}

\usepackage[round]{natbib}

\usepackage{amsthm}

\usepackage{graphicx}

\usepackage{microtype}

\usepackage[english]{babel}

\usepackage{bm}

\usepackage{enumitem}

\setenumerate{noitemsep,topsep=0pt,parsep=0pt,partopsep=0pt}
\setitemize{noitemsep,topsep=0pt,parsep=0pt,partopsep=0pt}

\usepackage{booktabs}

\usepackage{mathrsfs}

\usepackage{algorithm2e}

\usepackage{adjustbox}

\theoremstyle{definition}

\usepackage{mdframed}
\theoremstyle{plain}

\theoremstyle{definition}

\usepackage{ifxetex,ifluatex}
\usepackage{etoolbox}
\usepackage[svgnames]{xcolor}

\usepackage{tikz}

\usepackage{framed}

\usepackage{placeins}

\usepackage[right=1in, left=1in,top=1in, bottom=1in]{geometry}

\DeclarePairedDelimiter{\abs}{\lvert}{\rvert}

\usepackage{matlab-prettifier}
\makeatletter
\newcommand\incircbin
{%
	\mathpalette\@incircbin
}
\newcommand\@incircbin[2]
{%
	\mathbin%
	{%
		\ooalign{\hidewidth$#1#2$\hidewidth\crcr$#1\bigcirc$}%
	}%
}
\makeatother

\usepackage{extarrows}

\usepackage{shorttoc}

\DeclareMathOperator{\R}{\mathbb{R}}

\DeclarePairedDelimiter{\curl}{\lbrace}{\rbrace}

\mathtoolsset{showonlyrefs}
   
\usepackage{pdflscape}
\usepackage{longtable}

 \allowdisplaybreaks

\begin{document}

	\title{
		\vspace{-2cm} Is Productivity Advantage of Cities\\ Really  Down To Shift, Dilation, and Truncation?}

	\author{Vladislav Morozov \and Andrea Sy\vspace{-0.4cm} \thanks{ \scriptsize Mozorov: Independent. Email: vladislav.morozov@barcelonagse.eu ; Sy: CEMFI \& Banco de Espa\~na. Address: Calle de Casado del Alisal 5, 28014 Madrid, Spain. Email: andrea.sy@cemfi.es \\
The replication code for the paper is available at \url{https://github.com/symorz/productivity-cities-differences}.
\\
			 Acknowledgments: We are grateful to Julio Galvez for helpful comments. V. Morozov conducted this research while affiliated with the University of Bonn. The views expressed here are of the authors only, and not necessarily those of the Banco de Espa\~na or the Eurosystem.}	}

	\date{July 1, 2026}

\maketitle

\begin{abstract}

\noindent
Firms in denser areas are more productive, owing to agglomeration and selection. To disentangle these channels, \cite{Combes2012b} assume that total factor productivity (TFP) distributions in denser and less dense areas are identical up to shift, dilation, and truncation. Using Spanish firm-level data and methods robust to noisy TFP estimates, we find that TFP distributions are indeed statistically identical up to these parameters, validating such decompositions. Furthermore, shifts and dilations alone are sufficient to capture distributional differences, at least in Spain. This suggests that policymakers should focus on agglomeration policies.

\end{abstract}

\section{Introduction}

Firms in areas of higher population density are systematically more productive \citep{Melo2009, Combes2015EmpiricsAgglomerationEconomies, Duranton2020EconomicsUrbanDensity}. This productivity advantage may reflect agglomeration economies or firm selection --- the concentration of firms in dense environments making them more productive, or stronger competition in denser areas culling weaker firms at higher rates, respectively. Understanding which of these mechanisms drives the productivity advantage of cities is important for evaluating place-based interventions like tax incentives, infrastructure investments, or regional development programs \citep{Neumark2015PlaceBasedPolicies}.

\cite{Combes2012b} disentangle these channels by assuming that the distributions of total factor productivity (TFP) across denser and less dense areas are the same except for shift and dilation that reflect agglomeration and for tail truncation due to firm selection. Under this assumption, they decompose productivity differences along the two channels expressed by these three parameters. If, however, the TFP distributions across areas differed in other ways (e.g. tail behavior), this decomposition may be invalid and instead reflect those  differences, complicating interpretations of sorting patterns \citep{Gaubert2018FirmSortingAgglomeration} and evaluations of place‑based interventions.

Despite its central role, this assumption has not been directly tested. Empirical validation is challenging because productivity is not directly observed, but instead estimated with noise. Thus, naive comparisons of TFP distributions based on noisy TFP estimates may conflate (policy-irrelevant) noise with true heterogeneity \citep{Jochmans2019}.

This paper empirically tests this assumption of distribution equality. Using Spanish administrative firm-level data, we use a novel two-step nonparametric procedure to test for differences in TFP distributions across denser and less dense areas, while accounting for estimation noise in TFP. Across all sectors, we find that TFP distributions share the same shape, with differences fully explained by the parameters posited by \cite{Combes2012b}. 

Moreover, our results suggest that urban productivity advantages, at least in the context of Spain, may be fully explained by agglomeration rather than selection. Of the three \cite{Combes2012b} parameters, the shift and the dilation are sufficient to align the TFP distributions across areas, without relying on the tail truncation parameter. This suggests that policies enhancing agglomeration, such as improving infrastructure or fostering knowledge spillovers, may be more effective than competition-enhancing policies, such as reducing entry barriers or promoting firm turnover.

This paper makes two contributions. First, the key empirical contribution is to validate the distribution equality assumption in \cite{Combes2012b}. 
The GMM framework by \cite{Combes2012b} treats this equality as a maintained assumption to formulate its moment conditions, making the equivalence itself untestable within their framework. Our paper nonparametrically evaluates exactly the validity of this previously maintained assumption. It thus serves as validation for the use of the \cite{Combes2012b} decomposition and associated inference procedures. 
Second, our methodology contributes to the literature on working with noisy estimates of unobserved heterogeneity \citep{Okui2019b, Jochmans2019, Morozov2026InferenceExtremeQuantiles} by proposing a test for goodness-of-fit up to  transformations. Our two-step methodology for testing distribution equality with noisy data extends to other contexts, such as geographical differences in worker skills \citep{Roca2017}.

\section{Setting and Distribution Equality Hypotheses}\label{section:setting}
	
	To evaluate the assumption of distribution equality for TFPs, we work in the  setting of \cite{Combes2012b} (CDGPR12 henceforth). Their approach proceeds in two steps: first estimating firm-specific TFPs, then comparing productivity distributions between high-density (above-median density, AMD) and low-density (below-median density, BMD) areas.
	
	Following CDGPR12, we define firm-level TFP $\theta_i$ by assuming that firm $i$ in year $t$, sector $s$, and area $a\in\curl{\text{AMD}, \text{BMD}}$ produces value added $V_{i,t}$ according to a Cobb-Douglas production function:
	\begin{equation}\label{eq:cobb-douglas}
		V_{it} = \exp(\theta_i) K_{it}^{\beta_{1, s, a}} L_{i\, t}^{\beta_{2, s, a}}\exp(U_{it}+ \beta_{0, t, s, a})
	\end{equation}
	where $\theta_i$ is firm (log) total factor productivity (TFP), $K_{it}$ and $L_{it}$ are capital and labor, and $U_{it}$ captures measurement error in $V_{it}$. Here $\beta_{1, s, a}$ and $\beta_{2, s, a}$ are sector- and area-specific factor shares, while $\beta_{0, t, s, a}$ is a time-specific intercept, allowing for productivity trends. Firms in sector $s$ draw $\theta_i$ from distribution functions  $F_{s, AMD}$ (AMD) and $F_{s, BMD}$ (BMD).
	
	To separate agglomeration economies and selection, CDGPR12 assume that $F_{s, AMD}$ and $F_{s, BMD}$ are identical up to shift ($\mu$), dilation ($\sigma$), and left tail truncation $(\xi)$:
	\begin{equation}\label{eq:cdgpr_assm}
		H_0:\quad F_{s, AMD}(\theta) = \max\curl*{0, \dfrac{F_{s, BMD}\left( \frac{\theta-\mu}{\sigma} \right) - \xi}{ 1-\xi } } \text{ for all }\theta\in\R
	\end{equation}
	Here $(\mu, \sigma)$ capture agglomeration economies via shift and dilation, while $\xi$ reflects stronger competition in denser areas resulting in truncation of the left tail. CDGPR12 view \eqref{eq:cdgpr_assm} as a continuum of moments and estimate $(\mu, \sigma, \xi)$ with GMM.
	
	Our goal is to formally test the full assumption \eqref{eq:cdgpr_assm}. Because \eqref{eq:cdgpr_assm} yields the moment conditions for $(\mu, \sigma, \xi)$, it acts as a maintained assumption that cannot be evaluated from within the CDGPR12 framework itself. Yet, if the true TFP distributions differ in a way that deviates from \eqref{eq:cdgpr_assm}, the resulting estimates of agglomeration and selection may lose their structural interpretation and lead to erroneous conclusions about the drivers of urban productivity advantages and the effectiveness of place-based policies.
    
    Our paper thus takes a step back to test this maintained assumption. To do so validly and in a way that is compatible with the nonparametric spirit of CDGPR12 regarding the assumed common shape $F_{s, AMD}$ or $F_{s, BMD}$, our approach is agnostic economically and statistically:
\begin{itemize}[noitemsep,topsep=0pt,parsep=0pt,partopsep=0pt, leftmargin=*]
    \item  We do not require adopting any specific economic framework to explain how   $F_{s, AMD}$ or $F_{s, BMD}$ are formed. We test distributional equivalence directly, taking TFPs as primitives.
    \item We impose no parametric assumptions on the shape of $F_{s, AMD}$ or $F_{s, BMD}$.
\end{itemize}

\vspace{2mm}
	
	Testing the null hypothesis \eqref{eq:cdgpr_assm} directly is challenging for two main reasons. First, the TFPs $\theta_i$ are latent and must be estimated from data using some $\hat{\theta}_i$. As \cite{Jochmans2019} show, naively using these noisy estimates may lead to invalid inference about the underlying distributions. The most prominent source of this estimation noise in our context is the fact that firm-level TFPs are estimated from firm-level time series where the number of observations is quite limited (see section \ref{section:data}). Other sources of measurement error may also appear, such as differences due to using one or another production function estimator, the use of firm-level rather than establishment-level data, or classical measurement error in inputs.

    The second statistical challenge is that the hypothesis features both "regular" parameters ($\mu, \sigma$) and a tail truncation parameter ($\xi$). The "regular" parameters are estimable at a standard $\sqrt{N}$ rate both under the null and the alternative; they can be incorporated into uniform testing frameworks (like a Kolmogorov-Smirnov test). In contrast, for a test to have power against the alternative of \eqref{eq:cdgpr_assm},  $\xi$ has to reflect the tail of the distribution --- extreme quantiles (see appendix \ref{app:details}). Estimators related to extreme quantiles exhibit more exotic asymptotic behavior and need to  be treated with considerably more challenging specialized inference methods (see \cite{DeHaan2006} on general extreme value theory and \cite{Sasaki2022,Morozov2026InferenceExtremeQuantiles} on extreme value theory with noisy observations).
    
    To address these challenges, we propose a testing procedure in Section \ref{section:methodology} that handles both estimation noise and the differing statistical properties of these parameters.

\section{Data and TFP Estimation}\label{section:data}
 
We use administrative firm-level balance sheet and income statement data from the Banco de España's CBI dataset \citep[remote access, 2024 vintage]{BancoDeEspana2024MicrodataIndividualEnterprises} from 2000 to 2019. CBI is representative, covering  $\sim$80\% of incorporated non-financial Spanish firms and matching the aggregate dynamics of output, employment and wages \citep{Almunia2018}.  The dataset includes revenues, intermediate inputs, capital stock, employment, wage bill, sector identifiers, and postal codes, enabling us to construct firm-level TFP and classify firms by industry and urban area. All monetary variables are deflated using industry-specific deflators from EU-KLEMS.

Firms are assigned to AMD and BMD areas using the experienced density measure from \cite{Roca2017}, which averages population within a fixed radius around individuals to smooth boundary irregularities. Urban areas follow Spain’s Ministry of Housing definitions, excluding the exclaves Ceuta and Melilla. Firms are linked to urban areas via municipality-postal code concordances from the Instituto Nacional de Estadística.\footnote{In Spain, municipalities are administrative boundaries, while postal codes are operational areas that may cross municipal borders}

To address potential geographic misclassification for multi-establishment firms, we conservatively exclude firms with more than 50 employees. This threshold is motivated by evidence that firms below 100 employees are predominantly mono-establishment \citep{Xi2023_multi_estab_firms}.\footnote{\cite{Combes2012b} have establishment-level data but show their results are robust to including multi-establishment firms.} Finally, some additional standard data cleaning steps are detailed in Appendix \ref{sec.A1}.
 
Productivity estimates $\hat{\theta}_i$ are obtained in two steps, mirroring \cite{Combes2012b}:
\begin{enumerate}[noitemsep,topsep=0pt,parsep=0pt,partopsep=0pt, label={(\arabic*)}, leftmargin=*]
	\item For each sector $s$ and area $a$ (AMD/BMD), we estimate a gross output production function using the control functionapproach of \cite{DeLoeckerWarzynski2012} (in the spirit of \cite{Ackerberg2015};see Appendix \ref{sec.A2}).
	
	\item We compute $\hat{\theta}_i$ as the average residual from the estimated gross output production function:
	\begin{equation} \label{eq:tfp_est}
		\hat{\theta}_{i}= T^{-1}\sum_{t=1}^T \left[\log (Y_{it}) - \hat{\beta}_{0, s, t} -\hat{\beta}_{1, s} \log( K_{it}) - \hat{\beta}_{2, s} \log (L_{it}) -\hat{\beta}_{3, s} \log( M_{it}) \right]	,
	\end{equation}
    where $Y_{it}$ is gross output and $M_{it}$ is intermediate inputs.
    Following \cite{DeLoeckerWarzynski2012}, we use a more flexible gross output approach to allow materials to enter production freely, avoiding the fixed-proportions assumption implicit in value-added specifications.
\end{enumerate}

Importantly, for testing \eqref{eq:our_null}, we retain only firms observed for at least 15 periods to test the null \eqref{eq:our_null}. This serves two purposes:

\begin{enumerate}[noitemsep,topsep=0pt,parsep=0pt,partopsep=0pt, label={(\arabic*)}, leftmargin=*]
    \item Economically, it aligns our sample with the longer-run theoretical framework in \cite{Combes2012b} in which selection effects are expected to manifest (lemma 1). A cross-section of \textit{all} firms may include unproductive recent entrants not yet culled by competition, obscuring selection; such firms are less present in longer-lived samples. 
     
    \item Statistically, this restriction enables valid inference by sufficiently controlling the estimation noise in the TFP estimators $\hat{\theta}$, see section \ref{section:methodology} and the Appendix.
\end{enumerate}
The number of such firms varies between $149$ and $31433$, depending on the sector and area type (see table \ref{table:results} for exact counts). The 15 years threshold balances having long enough time series to capture selection, control estimation error (in the sense discussed in section \ref{section:methodology}) and large enough cross-sectional sample sizes for testing.

\section{Testing Distribution Equality}\label{section:methodology}

To test the equality assumption \eqref{eq:cdgpr_assm} while handling noisy TFP estimates and the distinct properties of estimators for shift, dilation, and truncation, we propose a two-step test that first checks a tighter sufficient condition and only proceeds to the full null if necessary.

\paragraph{Step 1: Test equality up to shift and dilation.} First, we test a tighter null hypothesis that matches distributions using only shift ($\mu$) and dilation ($\sigma$). Specifically, we test:
\begin{equation}\label{eq:our_null}
    H_{0, \text{step }1}:\quad F_{s, AMD}\left( \dfrac{\theta-\mu_{s, AMD}}{\sigma_{s, AMD}} \right) = F_{s, BMD}\left(\frac{\theta-\mu_{s, BMD}}{\sigma_{s, BMD} } \right) \text{  for all }\theta\in\R
\end{equation}
where $\mu$ and $\sigma$ are sector- and area-specific location and scale parameters. Based on the result of this test, two possibilities open up:
\begin{itemize}[noitemsep,topsep=0pt,parsep=0pt,partopsep=0pt, leftmargin=*]
    \item \textit{No reject:} If the step 1 null is not rejected, we conclude that the CDGPR12 assumption \eqref{eq:cdgpr_assm} indeed holds. The reason is that the step 1 null is a special case of the full CDGPR12 assumption \eqref{eq:cdgpr_assm} with $\xi=0$. Because allowing for a truncation parameter --- an additional degree of freedom --- can only improve the mapping between two distributions, distribution equality without truncation guarantees equivalence with it. The procedure concludes here.
    \item \textit{Reject:} If the step 1 null is rejected, we proceed to step 2.
\end{itemize}

\paragraph{Step 2: Test for tail truncation.} If step 1 rejects, we fully allow for selection (left-tail truncation) and test the full null \eqref{eq:cdgpr_assm}. If this re-test fails to reject, the data supports the full CDGPR12 assumption. Otherwise the null is rejected, indicating that the distributions differ beyond shift, dilation, and truncation. 

\paragraph{Motivation for approach}

This sequential approach is driven by information considerations: simply put, not rejecting in step 1 is a stronger signal that CDGPR12 equality \eqref{eq:cdgpr_assm} is true than not rejecting in step 2. There are two reasons for this. First, we again highlight that the null of step 1 is a possible (stronger) form of the full CDGPR12 assumption. 
Second, the step 1 test is better behaved --- relying on standard asymptotic theory and being straightforward to debias --- while step 2 requires estimating tail behavior for truncation, which introduces additional noise and non-standard convergence rates (see Appendix \ref{app:details}). Thus, when the step 1 null cannot be rejected, we place greater confidence in the conclusion.

\paragraph{Implementation}

In step 1, we use  a tailored nonparametric two-sample Kolmogorov-Smirnov (KS) test, modified to handle noisy TFP estimates and unknown parameters:

\begin{enumerate}[noitemsep,topsep=0pt,parsep=0pt,partopsep=0pt, label={(\arabic*)}, leftmargin=*]
    \item Half-panel jackknife (HPJ) debiasing \citep{Dhaene2014,Jochmans2019} to correct the estimation noise bias introduced by estimating firm-level TFP from short firm-level time series (using $\hat{\theta}_i$ instead of $\theta_i$).
    \item Firm-level bootstrap to account for all sources of sampling variability.
\end{enumerate}
The step 2 test is analogous but adjusts for truncation and uses subsampling (instead of bootstrapping) to handle the non-standard asymptotics of extreme quantiles.

The debiased test remains statistically valid even with noisy TFP estimates due to the restriction on the minimal number of observations per firm (see section \ref{section:data} and the Appendix).

The HPJ debiasing directly addresses the estimation noise bias. Furthermore, this correction is applied to estimators based on estimated TFPs and so it is agnostic to the specific production function estimator employed. The procedure may also reduce non-linearity biases arising from the first-stage production function estimation. We note that this debiasing relies on the underlying estimators being consistent as the time dimension grows. As such, it cannot compensate for systematic measurement error in inputs which would invalidate the production function estimates.

\section{Results and Discussion}\label{section:results}

\begin{figure}[!ht]
    \centering
    \includegraphics[width=\textwidth]{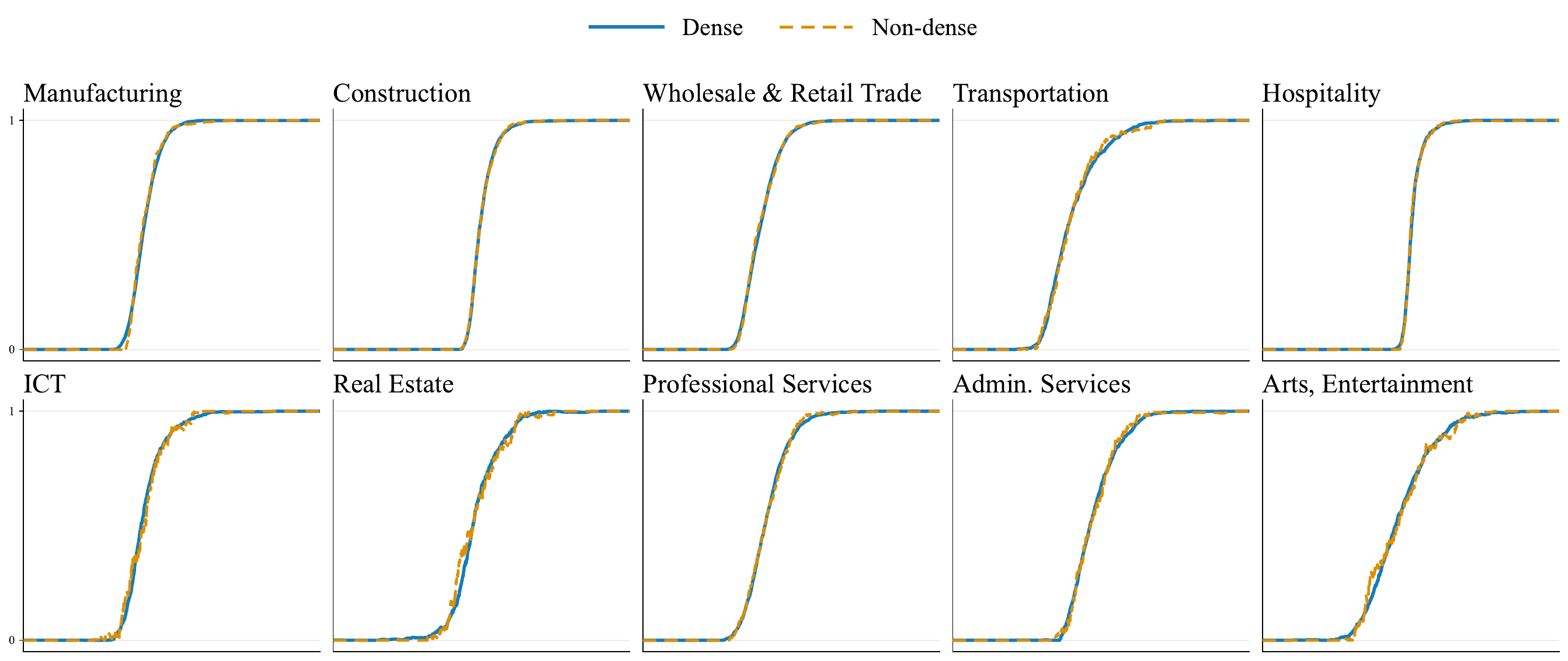}
    \caption{Comparison of debiased cumulative distribution functions (CDFs) of TFP in AMD and BMD areas. Adjusted for shift and dilation in each area and sector. Note: HPJ debiasing can make debiased CDF estimates nonmonotonic at some points \citep{Jochmans2019}}\label{fig:cdfplot}
\end{figure}

Table \ref{table:results} summarizes the results of step 1 of our testing procedure. It provides the $p$-values of our bias-corrected KS test for \eqref{eq:our_null} for each sector, along with the firm counts in AMD and BMD areas. Additionally, figure \ref{fig:cdfplot} depicts the debiased cumulative distribution functions (CDFs) of TFP. In line with the step 1 null \eqref{eq:our_null}, we standardize the CDFs to adjust for shift and dilation in both areas to allow for comparison of differences beyond these parameters (truncation or shape differences).

\begin{table}[!ht]
	\centering
	\begin{tabular}{|c|c|c|c|}
		\hline
		\textbf{Sector} & \textbf{$p$-value} & $N_{s, AMD}$ & $N_{s, BMD}$ \\
		 \hline\hline
		Manufacturing & 0.745 & 16374 & 1796 \\ \hline
		Construction & 0.991 & 10368 & 1826 \\ \hline
		Wholesale and Retail Trade & 0.686 & 31433 & 4635 \\ \hline
		Transportation & 0.821 & 3688 & 473 \\ \hline
		Hospitality & 0.643 & 6122 & 1180 \\ \hline
		ICT & 0.824 & 2434 & 171 \\ \hline
		Real Estate & 0.664 & 1146 & 149 \\ \hline
		Professional Services & 0.976 & 6524 & 688 \\ \hline
		Admin. Services & 0.974 & 2779 & 345 \\ \hline
		Arts, Entertainment & 0.634 & 1388 & 198 \\ \hline
		 \hline
	\end{tabular}
	\caption{Bootstrap $p$-values and sample sizes for testing Step 1 null \eqref{eq:our_null} of distribution equality up to shift and dilation, split by sector}\label{table:results}
\end{table}

Our results provide strong empirical support for the CDGPR12 distribution equality assumption \eqref{eq:cdgpr_assm}. For every sector, we fail to reject the null hypothesis that TFP distributions differ  between AMD and BMD areas only in the CDGPR12 parameters. Moreover, we reach this conclusion already in step 1 of our approach --- which is sufficient to conclude that the full equality assumption holds --- and hence step 2 is not triggered.  All $p$-values are large, with the smallest equal to 0.634. This pattern holds for both sectors with lower values of $N$ (e.g., Arts) and more populous ones (such as Trade). Visually, the shift-and-dilation standardized distribution functions in Figure \ref{fig:cdfplot} are extremely similar, particularly in more populous sectors where the estimates are more stable.

Interpreted narrowly, this finding validates the core assumption of CDGPR12 in the data, justifying their decomposition framework (and related methods) in prior work.

More broadly, step 1 already being sufficient suggests an even stronger relationship between TFP distributions: just two parameters --- shift and dilation --- suffice to align the distribution, at least in the context of our Spanish data. This suggests that the productivity differences between denser and less dense areas may be driven predominantly by mechanisms that operate through these channels, such as various aspects of  agglomeration economies.

These findings narrow the range of plausible explanations for productivity differences between areas, at least in the context of Spain. Since agglomeration economies are the leading explanation for shift and dilation, future research may focus on mechanisms operating through these channels. Policies aimed at fostering agglomeration (improvements in local infrastructure, labor market thickness, or knowledge spillovers) may help reduce productivity gaps between dense and sparse areas. Conversely, policies targeting firm selection (e.g., through competition policy) may be less effective in this context, given the high match attained with just shift and dilations.\footnote{
In settings with larger differences in market access, such as international comparisons, selection forces may be stronger, and truncation may arise \citep{Combes2012b}. Our methodology is designed to detect such patterns.}

Finally, we remark that our results are robust in two senses:

\begin{itemize}[noitemsep,topsep=0pt,parsep=0pt,partopsep=0pt, leftmargin=*]
    \item To trends in productivity and functional form assumptions. Our results are nonparametric in nature, requiring only correct specification of sectoral production functions. They allow for trends in productivity, impose no restrictions on the interpretation of shift-dilation parameters, and make no assumptions on functional forms of the TFP distributions.

\item To alternative production function estimators. Appendix \ref{sec.C} shows that using \cite{Olley1996}, \cite{Levinsohn2003}, \cite{Wooldridge2009}, or estimating our baseline approach at the sector level rather than sector-area level yields the same qualitative conclusion of not rejecting the null in step 1.

\end{itemize}

\section{Concluding Remarks}

In this paper, we empirically validate the key assumption of \cite{Combes2012b} that TFP distributions differ between denser and less dense areas only via shift, dilation, and tail truncation. Using Spanish firm-level data and econometric methods robust to noisy productivity estimates, we adopt a sequential testing approach. We find that this assumption holds across all sectors. Moreover,  shift and dilation alone are completely sufficient to account for differences in distributions within our domestic sample. Our results support the use of models relying on these parameters in policy analyses of urban productivity gaps and the design of place-based interventions. 

Our findings suggest two further areas of research. First, future work could explore the specific mechanisms driving shift and dilation in productivity distributions, such as local infrastructure, labor market thickness, or knowledge spillovers, to help policymakers design more effective place-based interventions. Second, our statistical approach --- combining HPJ debiasing with nonparametric distribution tests --- may also be used in other contexts where unobserved variables must be estimated with noise (e.g., worker skills in \cite{Roca2017} or mutual fund manager skills in \cite{Barras2022SkillScaleValuea}). Such extensions include testing distribution equivalence in international contexts, where selection effects may be  stronger \citep{Combes2012b}.

\begin{small}
	\setlength{\bibsep}{0pt plus 0.3ex} 
	\bibliographystyle{bbrr.bst}
	
	\bibliography{eco.bib}
	
\end{small}

\appendix
\centerline{\LARGE{\textbf{Appendix}} }

\counterwithin{figure}{section}
\setcounter{footnote}{0}        
\setcounter{section}{0}     
\renewcommand{\thetable}{\Alph{section}\arabic{table}}
\setcounter{figure}{0}      

\section{Further Details on Data and TFP Estimation}\label{sec.A}

\subsection{Data Preparation Details}\label{sec.A1}

As mentioned in section \ref{section:data}, we apply several cleaning steps to CBI data prior to production function estimation, as follows:

\begin{enumerate}[noitemsep,topsep=0pt,parsep=0pt,partopsep=0pt, label={(\arabic*)}, leftmargin=*]
	\item For years with missing data, we use the next year's values if available, following \cite{Gonzalez_Sy2025}. If both current and next year's filings exist, we prioritize the current year's values unless flagged as low quality.  
	\item We keep firms that have a legal status of corporation or limited liability company. This is inferred from the Spanish tax identification number of firms (\textit{CIF}/\textit{NIF}). We treat observations that switch legal status as observations of the same firm.
	\item For firms with multiple industry codes, we use the most frequent one.
	\item  We drop observations with negative or missing values for assets, revenues, tangible fixed assets, inputs, employees, or wage bill (following \cite{Orbis2024}).
	\item We exclude the following sectors, following \cite{Almunia2018}: financial services, public administration and defense, education, health and social work, and membership activities (NACE codes K, O, P, Q, and S). We also drop observations without sector classification or postal code information.
	\item We drop observations flagged as low quality. 
\end{enumerate}

\noindent
We also drop two-digit sectors with fewer than 300 observations in either AMD or BMD areas to maintain sample balance in estimation. After all cleaning steps, the final sample sizes for production function estimation range from 394 to 719,153 observations, depending on the sector and area. 

In the estimation of the production function following \cite{DeLoeckerWarzynski2012}, we measure gross output $Y_{it}$ using revenues; capital $K_{it}$ using tangible fixed assets; materials $M_{it}$ using material costs, defined as the difference between total inputs and other operating costs; and labor input $L_{it}$ using wages instead of employment, in order to account for differences in labor quality and actual hours worked across firms \citep{HsiehKlenow2009}. We deflate capital using industry‑specific capital deflators and deflate all other nominal variables using industry‑specific value‑added deflators from EU‑KLEMS. For the value-added production function estimators used in Appendix \ref{sec.C}, value-added $V_{it}$ is defined as the difference between revenues and material costs. All variables entering the production function estimation are expressed in logs.

\subsection{Production Function Estimation}\label{sec.A2}

As in \citet{DeLoeckerWarzynski2012}, we follow the control function approach of \citet{Ackerberg2015}, which builds on the methods of \citet{Olley1996} and \citet{Levinsohn2003} and addresses the simultaneity problem arising from the fact that firms observe their own productivity when choosing inputs.

Productivity $\theta_{it}$ can be expressed as a function of the firm's inputs:
\[
    \theta_{it} = h_t(k_{it}, l_{it}, m_{it}),
\]
where $k_{it}$, $l_{it}$ and $m_{it}$ are $\log$ of capital, labor and materials respectively. 

We follow \cite{Ackerberg2015} and impose a timing structure on input choices. Capital is predetermined and fixed at the beginning of period $t$. Labor is assumed to be chosen at $t-1$, reflecting hiring and firing costs or contractual rigidities that make labor adjustment sluggish.\footnote{\citet{Ackerberg2015} treat predetermined labor as a special case within their general timing framework, which allows labor to be chosen either within the period or in advance.}  Materials, by contrast, are assumed to be a fully flexible input and used as a proxy for productivity.

The estimation proceeds in two steps. In the first step, we obtain an estimate of expected output net of measurement error and unanticipated shocks:
\[
    y_{it} = \beta_0 + \beta_k k_{it} + \beta_l l_{it} + \beta_m m_{it} + h_t(k_{it}, l_{it}, m_{it}) + \varepsilon_{it} \equiv \Phi_t(k_{it}, l_{it}, m_{it}) + \varepsilon_{it} 
\]
To approximate $\Phi_t(\cdot)$, we run an OLS regression of $y_{it}$ on a third-order polynomial in $(k_{it}, l_{it}, m_{it})$ and obtain the estimate $\widehat{\Phi_t}$.

In the second step, productivity is assumed to follow a first-order Markov process:
\[
    \theta_{it} = g_t(\theta_{i,t-1}) + \xi_{it},
\]
where $g(\cdot)$ is approximated using a third-order polynomial and $\xi_{it}$ is the innovation in productivity -- the part that firms cannot predict when choosing inputs.

Using the estimated productivities, we estimate the function $g(\cdot)$ by regressing $\omega_{it}(\beta)$ on a polynomial in $\omega_{it-1}(\beta)$. This gives us a fitted value $\widehat{g}(\omega_{it-1})$ and a residual
\[
        \widehat{\xi}_{it}(\beta) = \omega_{it}(\beta) - \widehat{g}(\omega_{it-1})
 \]

Based on these steps, the production function parameters are identified using the GMM moment conditions:
\[
     \mathbb{E}[\widehat{\xi}_{it} \times \mathbf{z_{it}} ]  = 0
\]
where the instrument vector is $\mathbf{z_{it}} = [1 \ m_{it-1} \ l_{it} \ l_{it-1} \ k_{it}]'$.

These instruments are predetermined with respect to period-$t$ productivity and address the timing critique of \citet{Ackerberg2015}. The production function is estimated separately by 2-digit sector--area (AMD/BMD).


\section{Technical Details}\label{sec.B}

\paragraph{Test Statistic for Step 1}

Formally, the test statistic for each sector $s$ is given by
\begin{equation}\label{eq:test_stat}
	T_s = \sup_{\theta\in\R} \abs*{ \check{F}_{s, AM}\left( \dfrac{\theta-\check{\mu}_{s, AM}}{\check{\sigma}_{s, AM}} \right) - \check{F}_{s, BM}\left(\frac{\theta-\check{\mu}_{s, BM}}{\check{\sigma}_{s, BM} } \right)    },
\end{equation}
where $\check{F}_{s, a}$, $\check{\mu}_{s, a}$, and $\check{\sigma}_{s, a}$ are the HPJ-debiased estimators of  CDF, mean, and variance constructed based on the empirical CDF and sample mean and variance for $a\in\curl{\text{AMD}, \text{BMD}}$.

\paragraph{HPJ Debiasing}
   
The key challenge is that $\hat{\theta}_i$ --- our TFP estimators  --- are noisy measurements of $\theta_i$. As \cite{Jochmans2019} show, using $\hat{\theta}_i$ instead of $\theta_i$ causes an $O(1/T)$ bias in the  estimated CDF ($T$ is the number of observations per firm). HPJ debiasing \citep{Dhaene2014} reduces this bias to $O(1/T^2)$ by:
\begin{enumerate}[noitemsep,topsep=0pt,parsep=0pt,partopsep=0pt, label={(\arabic*)}, leftmargin=*]
	\item Splitting the panel into two contiguous time-based halves.
	\item Computing $\hat{\theta}_i$ separately on each half still using formula \eqref{eq:tfp_est}.
	\item Constructing estimators $\hat{\phi}_{1, s, a}$ and $\hat{\phi}_{2, s, a}$ for $\phi\in\curl{\mu_{s, a}, \sigma_{s, a}, F_{s, a}}$ based on samples of $\hat{\theta}_i$  from the first and second subpanels, respectively.
	\item Computing full panel estimators $\hat{\phi}_{s, a}$ based on $\hat{\theta}_i$ that use the full dataset.
	\item Constructing debiased estimators $\check{\phi}_{s, a}$ as $
		\check{\phi}_{s, a} = 2\hat{\phi}_{s, a} - \frac{1}{2}\left( \hat{\phi}_{1, s, a} + \hat{\phi}_{2, s, a} \right)$.
	
\end{enumerate}  

\paragraph{Bootstrap Inference}

To compute the critical values for $T_s$, we use firm-level bootstrap. In each iteration, we resample firms with replacement and recompute $\mu_{s, a}, \sigma_{s, a}$, and $F_{s, a}$ with HPJ debiasing. We tabulate the bootstrap values of the test statistic \eqref{eq:test_stat}. Finally, the actual value of $T_s$ is compared to the bootstrap distribution to obtain the $p$-value of the test. 

\paragraph{Validity}

The test's validity follows from \cite{Jochmans2019} (section 3.1) and theory for KS tests with estimated parameters \citep{Durbin1975KolmogorovSmirnovTestsWhen, Khamis1992DCorrectedKolmogorovSmirnovTest}. Debiasing is key. Without it, validity would require the stricter condition $\max\curl{N_{s, AMD}, N_{s, BMD}}/T^2\sim 0$. With HPJ, the weaker condition  $\max\curl{N_{s, AMD}, N_{s, BMD}}/T^4\sim 0$ suffices, which holds in our data by construction (see section \ref{section:data} and table \ref{table:results}).

\paragraph{Details for Step 2}\label{app:details}

Step 2 is analogous to Step 1, with two differences. First, the test statistic is analogous to \eqref{eq:test_stat} but with the CDFs also adjusted for the estimated truncation parameter $\hat{\xi}$. Second, inference relies on subsampling instead of the bootstrap. 

To estimate the parameters $(\mu, \sigma, \xi)$,  we minimize quantile distance between empirical distributions of the samples using the approach of \cite{Freitag2005} and \cite{Gobillon2010}. To ensure that the test has power against the alternative of \eqref{eq:cdgpr_assm},  distance is minimized over a growing space of quantiles that asymptotically covers the entire support.

Including $\xi$ leads to several  complications relative to step 1: 

\begin{itemize}[noitemsep,topsep=0pt,parsep=0pt,partopsep=0pt, leftmargin=*]

      \item \textit{Dependence on extreme quantiles}: to have power against the alternative, the test must minimize the distance between the two distributions across growing support, including  tails. The distance-minimizing estimators of $(\mu, \sigma, \xi)$ hence need to rely on extreme quantiles.
       
      Extreme quantiles lead to three challenges: different properties under estimation noise \citep{Morozov2026InferenceExtremeQuantiles}, convergence rates that differ from $\sqrt{N}$ \citep{DeHaan2006}, and failure of standard bootstrap for inference \citep{Politis1999}.
    \item \textit{Non-uniformity:} If empirical selection is weak or absent, the truncation parameter $\xi$ may not be strictly bounded away from zero. If $\xi\approx 0$, the  $\sqrt{N}$  rates derived by \cite{Gobillon2010} for quantile minimum distance estimators  may break down (see e.g. \cite{Andrews2000InconsistencyBootstrapWhen} regarding such non-uniformity).
  
\end{itemize}
 To overcome these challenges, Step 2 inference instead relies on subsampling, with conditions for validity stemming from \cite{Jochmans2019} and \cite{Morozov2026InferenceExtremeQuantiles}.

\section{Robustness to Alternative TFP Estimation Methods}\label{sec.C}
\setcounter{table}{0}
\setcounter{figure}{0}

We assess the robustness of our distributional test to alternative production function estimators. We construct TFP using three standard value‑added estimators: \citet{Olley1996} (OP), which use investment as a proxy for unobserved productivity; \citet{Levinsohn2003} (LP), which instead rely on materials as the proxy; and \citet{Wooldridge2009}, which rewrites the LP moment conditions to avoid the first‑stage identification failure of OP/LP highlighted by \citet{Ackerberg2015} and estimates the model in a single GMM step. We also estimate TFP following \citet{DeLoeckerWarzynski2012} at the sector level to mirror the sector‑by‑sector approach in \citet{Combes2012b}.

Table \ref{table:robust_pval} reports sector‑level $p$‑values. The qualitative pattern of $p$‑values is highly stable across estimators and across the two levels of TFP estimation (sector vs. sector–area). This shows that our results are not driven by the specific control function approach used to address input endogeneity, nor by the additional geographic granularity in our baseline TFP estimates.

\begin{table}[!ht]
\centering
\caption{Bootstrap $p$-values for testing Step 1 null \eqref{eq:our_null} of distribution equality up to shift and dilation, split by sector, across production function estimators.}
\label{table:robust_pval}
\begin{tabular}{lccccc}
\toprule
\textbf{Sector} & 
\textbf{OP} & 
\textbf{LP} & 
\textbf{Wooldridge} & 
\textbf{DLW$_s$} & 
\textbf{DLW$_{sa}$} \\
\midrule
Manufacturing            	& 0.924 & 0.976 & 0.856 & 0.902 & 0.745 \\
Construction             	& 0.583 & 0.612 & 0.582 & 0.845 & 0.991 \\
Wholesale and Retail Trade  & 0.788 & 0.759 & 0.707 & 0.630 & 0.686 \\
Transportation           	& 0.944 & 0.750 & 0.861 & 0.972 & 0.821 \\
Hospitality              	& 0.680 & 0.892 & 0.907 & 0.824 & 0.643 \\
ICT                      	& 0.679 & 0.865 & 0.663 & 0.627 & 0.824 \\
Real Estate              	& 0.798 & 0.709 & 0.709 & 0.549 & 0.664 \\
Professional Services    	& 0.785 & 0.862 & 0.920 & 0.999 & 0.976 \\
Admin. Services         	& 0.592 & 0.553 & 0.791 & 0.859 & 0.974 \\
Arts, Entertainment    		& 0.934 & 0.992 & 0.981 & 0.798 & 0.634 \\
\bottomrule
\end{tabular}
\begin{minipage}{\textwidth}
\vspace{2mm}
\small{\textit{Notes:} OP = \citet{Olley1996}; LP = \citet{Levinsohn2003}; Wooldridge = \citet{Wooldridge2009}; DLW = \citet{DeLoeckerWarzynski2012}. DLW$_s$ refers to DLW estimated separately for each sector while DLW$_{sa}$ corresponds to the to the baseline results in the main text (see Table \ref{table:results}), where production functions are estimated separately for each sector--area pair. } 
\end{minipage}
\end{table}

\end{document}